\documentclass[twocolumn]{aastex62}

\usepackage{amsmath}
\usepackage{lipsum}
\usepackage{float}
\usepackage{afterpage}
\usepackage{makecell}
\usepackage{upgreek}
\usepackage{bm}
\usepackage{ulem}

\newcommand{\Disco}{{\texttt{Disco}}}
\newcommand{\tensorSymbol}[1]{{\overset{\leftrightarrow}{#1}}}
\newcommand{\viscNew}{\text{visc}}
\newcommand{\viscOrig}{\text{visc \texttt{v1.0}}}

\graphicspath{{./}{figures/}}
\shorttitle{}
\shortauthors{Dittmann \& Ryan}

\begin{document}

\title{Preventing Anomalous Torques in Circumbinary Accretion Simulations}

\correspondingauthor{Alexander J. Dittmann}
\email{dittmann@astro.umd.edu}

\author[0000-0001-6157-6722]{Alexander J.~Dittmann}
\affil{Department of Astronomy and Joint Space-Science Institute, University of Maryland, College Park, MD 20742-2421, USA}
\affil{Center for Computational Astrophysics, Flatiron Institute, 162 5th Avenue, New York, NY 10010, USA}
\author[0000-0001-9068-7157]{Geoffrey Ryan}
\affil{Department of Astronomy and Joint Space-Science Institute, University of Maryland, College Park, MD 20742-2421, USA}
\affil{Astroparticle Physics Laboratory, NASA Goddard Space Flight Center, Greenbelt, MD 20771, USA}

\begin{abstract} 
Numerical experiments are the primary method of studying the evolution of circumbinary disks due to the strong nonlinearities involved. Many circumbinary simulations also require the use of numerical \emph{mass sinks}: source terms which prevent gas from unphysically accumulating around the simulated point masses by removing gas at a given rate.
However, special care must be taken when drawing physical conclusions from such simulations to ensure that results are not biased by numerical artifacts. 
We demonstrate how the use of improved sink methods reduces some of these potential biases in  vertically-integrated simulations of aspect ratio 0.1 accretion disks around binaries with mass ratios between 0.1 and 1. Specifically, we show that sink terms that do not reduce the angular momentum of gas relative to the accreting object: 1) reduce the dependence on the sink rate of physical quantities such as the torque on the binary, distribution of accretion between binary components, and evolution of the binary semi-major axis; 2) reduce the degree to which the sink rate affects the structure of the accretion disks around each binary component; 3) alter the inferred variability of accretion onto the binary, making it more regular in time.
We also investigate other potential sources of systematic error, such as the precise from of gravitational softening and previously employed simplifications to the viscous stress tensor. 
Because of the strong dependence of the orbital evolution of the binary on both the torque and the distribution of mass between binary components, the sink methods employed can have a significant effect on the inferred orbital evolution of the binary. 
\end{abstract}


\section{Introduction} 
Accretion disks surrounding binary systems are important to a broad range of astrophysical scenarios. Such circumbinary disks may occur as galaxies merge, potentially assisting the merger of supermassive black holes \citep[e.g.][]{{1980Natur.287..307B},{2000ApJ...532L..29G},{2009ApJ...700.1952H}}. Understanding the variability of accretion onto such binaries, and being able to disentangle it from standard variability in active galactic nuclei (AGN), is essential for accurately identifying binaries in photometric surveys \citep[e.g.][]{{2018ApJ...859L..12L},{2021MNRAS.500.4025L}}. Additionally, the circumbinary disk may leave imprints on intermediate and extreme mass ratio inspirals observed by LISA \citep{{2011PhRvL.107q1103Y},{2011PhRvD..84b4032K},{2019MNRAS.486.2754D},{2021MNRAS.501.3540D}}.
Stars may also form or be captured into AGN disks, forming compact objects, migrating, merging within the disk, and emitting gravitational waves detectable by LIGO \citep[e.g.][]{{1980SvAL....6..357K},{1993ApJ...409..592A},{2017MNRAS.464..946S}, {2020MNRAS.493.3732D},{2020ApJ...889...94M},{2020MNRAS.498.4088M},{2020arXiv200903936C}}. Circumbinary disks also naturally occur as a byproduct of binary star formation \citep[e.g.][]{{1986ApJS...62..519B},{2010ApJ...710.1375K}} and could be responsible for the population of widely-separated binaries with mass ratios $q\gtrsim0.95$ \citep{2019MNRAS.489.5822E}, and observed bimodality in the eccentricities of post-asymptotic giant branch stars \citep{{2018A&A...620A..85O},{2020arXiv201009707Z}}. 

A thorough understanding of circumbinary disks requires probing the wide parameter space of disk aspect ratios ($h/r$), binary mass ratios ($q$), orbital eccentricities ($e$), and inclinations ($i$), in addition to magneto-hydrodynamic turbulence, often approximated as a viscosity. Moreover, the nonlinearities present in the system limit the applicability of analytic theory and necessitate numerical experiments. One such example is the repeated finding of positive torque on the binary for some mass ratios and Mach numbers, driving the binary apart \citep[e.g.][]{{2017MNRAS.466.1170M},{2019ApJ...875...66M},{2019ApJ...871...84M},{2020ApJ...901...25D},{2020ApJ...889..114M}} rather than inward as predicted by analytic theory \citep{1991MNRAS.248..754P} and early simulations \citep{2008ApJ...672...83M}. However, simulating circumbinary disks along with the binary itself introduces other parameters to the problem, such as choices in gravitational softening and sinks to remove mass as it accretes onto each point mass. Sinks have been particularly controversial, as \citet{2017MNRAS.469.4258T} reported that by changing the rate at which mass is removed, the torque on the binary may not only change by orders of magnitude, but also change sign. \citet{2020ApJ...901...25D} later concluded, using higher-resolution simulations, that order of magnitude changes in the sink rate had only minor changes on the gravitational torque on the binary, but found that changes in sink rate could have significant effects on accretion variability and distribution of accreted mass between each binary component.

Sinks are small regions of a hydrodynamic simulation domain which absorb local gas mass, momentum, and energy.  They are a necessary component of many large scale accretion simulations, removing gas that would otherwise unphysically pile-up around gravitating masses due to finite numerical resolution.  Sinks are meant to approximately represent the local, rapid, and small-scale processes by which gas is ultimately accreted onto the gravitating objects below the scale of the numerical grid.  An effective sink prescription provides an accurate sub-grid model for accretion, allowing the simulation bulk to evolve as if the sink region were much more highly resolved.

By removing angular momentum, sinks can naturally cause a torque on local gas. In a steady state this will set up an angular momentum flux which can strongly effect the local surface density profile far from the accretor \citep{1974MNRAS.168..603L}. A good sink prescription should then only remove angular momentum appropriate for the object it represents.  In a thin Keplerian disk the angular momentum of a fluid element is proportional to the square root of the distance to the central accreting object. The amount of angular momentum accreted (i.e. the torque) is set by the specific angular momentum of material near  the ``surface'' of the object (the ISCO in the case of a black hole).  If the physical size of an accreting object in a simulation is much smaller than the numerical grid scale then the accreted angular momentum will necessarily be a minuscule fraction of the angular momentum in any nearby grid cell.  If a sink simply absorbs some fixed fraction of the local angular momentum, it will produce a large anomalous torque on the gas which can affect the global flow.  This issue has been discussed previously by \citet{2020ApJ...892L..29D}, who suggest sink particles that do not exert any torque on the gas (`torque-free' sinks), and investigate the effects of the sink method on steady-state disks. 

In the present work, we seek sink prescriptions that reduce the effect of numerical, unphysical, parameters on the evolution of accreting binaries. We apply the torque-free sink prescription presented in \citet{2020ApJ...892L..29D} to circumbinary disks in the range of mass ratios $0.1 \leq q \leq 1$ with fixed $h/r \approx 0.1$. We also study the influence of other numerical factors such as the functional form of gravitational smoothing, sink rates, and implementations of viscous terms. In general, we find that torque-free sinks reduce the degree to which sink rate affects the inferred evolution of binaries. 

\section{Methods}\label{methods}
The numerical experiments constituting this study used the moving-mesh code \Disco{} \citep{2016ApJS..226....2D} to solve the 2D equations of isothermal vertically-integrated hydrodynamics:
\begin{align}
\partial_t\Sigma + \bm{\nabla}\!\cdot\!(\Sigma\mathbf{v}) &= S_\Sigma \label{eq:continuity} \\
\partial_t(\Sigma\mathbf{v}) + \bm{\nabla}\!\cdot\!(\Sigma\mathbf{v}\mathbf{v}+P\tensorSymbol{I} - 2 \Sigma \nu \tensorSymbol{\sigma})
&= -\Sigma\bm{\nabla}\Phi + \mathbf{S}_{p} \label{eq:momentum} \\
P&=c_s^2(\mathbf{x})\Sigma,
\end{align}
where $P$ is the vertically-integrated pressure, $\Sigma$ is the disk surface density, $c_s$ is the sound speed, $\mathbf{v}$ is the fluid velocity vector, $\Phi$ is the gravitational potential, $\tensorSymbol{I}$ is the identity matrix, $S_\Sigma$ is a mass sink term, $S_{\Sigma\mathbf{v}}$ is a momentum sink term, $\tensorSymbol{\sigma}$ is the velocity shear tensor, and $\nu$ is the kinematic viscosity. The sink and source terms will be described later in detail.

The sound speed is computed based on the local gravitational potential, a variation of the commonly-adopted `locally isothermal' equation of state
\begin{equation}\label{lociso}
\begin{split}
c_s^2(\mathbf{x}) = -\Phi(\mathbf{x})/\mathcal{M}^2,
\end{split}
\end{equation}
where $\mathcal{M}$ is the Mach number. We chose this equation of state, and fix $\mathcal{M}=10$, to make comparisons with previous results in the literature \citep[e.g.][]{{2017MNRAS.466.1170M},{2019ApJ...871...84M},{2020ApJ...901...25D}}. We note that higher Mach numbers are expected in many astrophysical systems, and binary evolution at higher Mach numbers can be qualitatively different \citep{2020ApJ...900...43T}. Additionally, we note that locally-isothermal equations of state imply faster cooling timescales than may be applicable in many accretion disks \citep{2020ApJ...892...65M}. However, we target the present investigation to be more easily compared to others in the literature.

The simulations used piecewise linear reconstruction with a generalized van Leer slope limiter \citep{{1979JCoPh..32..101V},{2000JCoPh.160..241K}}, setting $\theta=1.5$, a Harten-Lax-Van Leer-Contact approximate Riemann solver \citep{1994ShWav...4...25T} and 2nd-order total variation diminishing Runge-Kutta time stepping \citep{1998MaCom..67...73G} with a Courant–Friedrichs–Lewy safety factor of 0.5. Each grid annulus was set to move with the $\phi$-averaged azimuthal fluid velocity in that annulus.

For the sake of numerical stability, we employ either of two softened gravitational potentials. In each case, the total potential is given by $\Phi=\sum_i \Phi_i$, where $\Phi_i$ is the gravitational potential of a point mass of mass $M_i$ at position $\mathbf{x}_i$.  For Plummer spheres, the potential from each body is 
\begin{equation}
\Phi_i = \frac{GM_i}{\sqrt{|\textbf{x}-\textbf{x}_i|^2+\epsilon_g^2}},
\end{equation}
where $\epsilon_g$ is the gravitational softening length.
We also consider spline softening \citep{2001NewA....6...79S}, where $\Phi_i = \frac{GM_i}{h}W(|\textbf{x}-\textbf{x}_i|/h)$, and 
\begin{equation}
W(u) = \begin{cases}
\frac{16}{3}u^2-\frac{48}{5}u^4+\frac{32}{5}u^5-\frac{14}{5}, & 0 \leq u < \frac{1}{2} \\
\frac{1}{15u}+\frac{32}{3}u^2-16u^3+\frac{48}{5}u^4\\ - \frac{32}{15}u^5 - \frac{16}{5}, & \frac{1}{2} \leq u < 1 \\
-\frac{1}{u}, & u\geq 1.
\end{cases}
\end{equation}
This formulation trades simplicity for the ability to exactly reproduce the Newtonian potential for $|\textbf{x}-\textbf{x}_i| \geq h$ However, both of these methods cause the equilibrium structure of thin accretion disks to deviate from the Newtonian solution on length scales similar to the smoothing length scale, as shown in Figure \ref{fig:singleDisks}. For consistency of notation, we set $h=\epsilon_g/2.8$ so that both potentials have the same minimum value for a given $\epsilon_g$.

\subsection{Sink Terms}

When point masses are included on a numerical grid, they naturally accumulate gas mass due to their gravity. Realistically this gas would be compressed onto the point mass at length scales far below the numerical grid scale.  However, in grid-based hydrodynamics codes such as \Disco{}, without explicit sub-grid treatment this gas will pile up over several grid cells and can eventually alter the gas dynamics of the global flow.

Often, sink terms are introduced, as in Equations (\ref{eq:continuity}) and (\ref{eq:momentum}) to ameliorate this problem. Ideally, such a sink term would prevent the unphysical buildup of material on the grid, but otherwise leave the gas dynamics unchanged by its presence. For example, when sink particles represent unresolved point masses, a sink term should preserve the physics of the system as close to the limit of infinite resolution as possible.

The surface density sink term in \Disco{} is given by 
\begin{equation}\label{surfSink}
S_\Sigma = -\gamma \Omega_b \Sigma \sum_i s_i(|\textbf{x}-\textbf{x}_i|),
\end{equation}
where $\Omega_b$ is the angular frequency of the binary, $s_i$ is a function specifying the sink profile for each particle, and $\gamma$ is the sink rate, which determines the rate at which material is removed. For our circumbinary simulations, we use 
\begin{equation}\label{GaussSquared}
s_i = \exp\left(-\frac{|\mathbf{x}-\textbf{x}_i|^4}{r_s^4}  \right),
\end{equation}
where $r_s$ is a characteristic sink radius. For simulations of accretion disks with single objects we use
\begin{equation}
s_i = \begin{cases}
(1-|\mathbf{x}-\textbf{x}_i|^2r_s^{-2})^2, & |\mathbf{x}-\textbf{x}_i| <r_s \\
0, & |\mathbf{x}-\textbf{x}_i| \geq r_s
\end{cases}
\end{equation}
and set $\Omega_b=1$ in Equation \eqref{surfSink}.

The treatment of the sink term in the momentum equation is more subtle. Traditionally \citep[e.g.][]{2019ApJ...871...84M,2020ApJ...900...43T, 2020ApJ...901...25D}, this term is implemented as
\begin{equation}\label{eq:standardMomSink}
    \mathbf{S}_{p}=S_\Sigma\mathbf{v}.
\end{equation}
However, as emphasized by \citet{2020ApJ...892L..29D}, such a sink term introduces a torque on gas as it acts. \citet{2020ApJ...892L..29D} demonstrated that such sinks can change the steady-state surface density of accretion disks in a way that depends on the value of $\gamma$, even well outside of $r_s$. This is undesired behavior, as $\gamma$ is a numerical tuning parameter which would ideally not change the bulk properties of the flow.  \citet{2020ApJ...892L..29D} introduced `torque-free' sinks, such that the action of the sink term does not change the angular momentum of accreting gas. This is useful because stellar surfaces and innermost stable orbits are rarely resolved in global simulations. Thus, gas would realistically need to shed angular momentum before it could accrete, and Equation \eqref{eq:standardMomSink} vastly overestimates the angular momentum removed from the fluid.

We implement torque-controlled sinks in \Disco{} using
\begin{align} 
    \mathbf{S}_{p} &= -\gamma \Omega_b \Sigma \sum_i s_i(|\textbf{x}-\textbf{x}_i|)\mathbf{v}^*_i \label{eq:tf1} \\
    \mathbf{v}^*_i &= \left(\mathbf{v} - \mathbf{v}_i\right)\cdot \left(\hat{\mathbf{r}}_i \hat{\mathbf{r}}_i + \delta\ \! \hat{\bm{\phi}}_i \hat{\bm{\phi}}_i \right) + \mathbf{v}_i, \label{eq:tf2}
\end{align}
where $\mathbf{v}_i$ is the velocity of the sink particle,  $\hat{\mathbf{r}}_i$ and $\hat{\bm{\phi}}_i$ are the unit basis vectors at position $\mathbf{x}$ of a polar coordinate system centered on the sink particle, and $\delta$ is a dimensionless control parameter.  The sink velocity $\mathbf{v_i^*}$ is then the gas velocity $\mathbf{v}$ corrected to have reduced angular velocity in the frame of the moving sink particle, controlled by the parameter $\delta$: if $\delta=0$ the sink becomes torque-free, while if $\delta=1$ Equations \eqref{eq:tf1} and \eqref{eq:tf2} reduce to Equation \eqref{eq:standardMomSink} and the ``standard'' prescription is recovered. A small nonzero $\delta$ could be used to model a small degree of spin-up when modeling objects the physical size of which is not too far below the grid scale, although we consider only the `standard' ($\delta=1$) and `torque-free' ($\delta=0$) limits in this work. The torque-controlled sinks may be easily extended to three dimensions by adding a $\delta\hat{\bm{\theta}}_i\hat{\bm{\theta}}_i$ term to Equation \eqref{eq:tf2}.

\subsection{Diagnostics}

The total mass accretion rate onto a sink particle $\dot{M}_i$ is given by the negative integral of the corresponding sink term over the entire two dimensional simulation domain:
\begin{equation}
    \dot{M}_i = - \int\! dA\ S_{\Sigma, i}\ ,
\end{equation}
where $S_{\Sigma, i} = -\gamma \Omega_b \Sigma s_i(|\mathbf{x}-\mathbf{x}_i|)$.  We can similarly define $\mathbf{S}_{p, i} = S_{\Sigma_i} \mathbf{v}_i^*$ by Equation \eqref{eq:tf1}.  Although the integral is formally taken over the entire domain, the sink profile function $s_i(x)$ sharply truncates the contributions of any region more than a few $r_s$ distance from the sink particle. 

The \emph{accretion torque} $\dot{J}_a$ is the rate of angular momentum delivered through the sink terms.  The accretion torque on a single sink particle is:
\begin{align}
    \dot{J}_{a,i} &= - \int \! dA\ \mathbf{x} \times \mathbf{S}_{p, i} = - \int \! dA\ S_{\Sigma, i}\ \mathbf{x} \times \mathbf{v}^*_{i}\ .
\end{align}
The accretion torque contains three independent contributions, corresponding to the three terms in $\mathbf{v}^*_i$.  The first is proportional to $\hat{\mathbf{r}}_i$ and corresponds to the direct absorption of linear momentum from the gas, the second is proportional to $\hat{\bm{\phi}}_i$ and contributes dominantly to the spin of the sink particle, and the third is proportional to the sink particle velocity $\mathbf{v}_i$ and corresponds to the direct accretion of mass to the sink.

The \emph{spin torque} $\dot{J}_{a, s}$ is the component of $\dot{J}_a$ responsible for increasing the internal angular momentum of the sink and does not contribute to the orbital angular momentum.  It is most directly computed in the instantaneous rest-frame of the sink particle:
\begin{align}
    \dot{J}_{a,s,i} &= - \int \! dA\ \left(\mathbf{x} - \mathbf{x}_i\right) \times \left(\mathbf{S}_{p, i} - \mathbf{v}_i S_{\Sigma, i} \right)\\
    &= - \int \! dA\ \delta \ S_{\Sigma, i}\left|\mathbf{x} - \mathbf{x}_i\right| \left(\mathbf{v} - \mathbf{v}_i\right)_{\hat{\phi}_i} \label{eq:spinTorque}
\end{align}

We can see that indeed, when $\delta = 0$ the sink particle experiences no spin torque and the sink is ``torque-free''.  The accreted orbital torque $\dot{J}_{a, o}$ is simply the difference of the total accreted torque and the spin torque:
\begin{equation}
    \dot{J}_a = \dot{J}_{a, o} + \dot{J}_{a, s}.
\end{equation}

Finally, sink particles may acquire orbital momentum through gravitational interactions. These are also simply computed via integrating the corresponding source terms over the simulation domain:

\begin{equation}
    \dot{J}_{g,i} =  -\int \! dA\ \mathbf{x} \times \left(-\Sigma \bm{\nabla} \Phi_i\right) =  \int \! dA\ \Sigma \partial_\phi \Phi_i
\end{equation}

The total torque on the system of sink particles is the sum of the torques on each component:
\begin{equation}
    \dot{J} = \sum_i \dot{J}_{g, i} + \dot{J}_{a,i} .
\end{equation}
The rate of change of the total \emph{orbital} angular momentum, the orbital torque $\dot{J}_{\mathrm{orb}}$, is: 
\begin{equation}
    \dot{J}_{\mathrm{orb}} = \sum_i \dot{J}_{g, i} + \dot{J}_{a,o,i} .
\end{equation}

We record these torques in \Disco{} as well as the accretion rate onto each object by tracking the source terms each time they are evaluated, and regularly storing the integrated mass and angular momentum of each sink particle in an output file along with the simulation time, and then resetting the dummy variables used for storage. This method effectively time averages over every hydrodynamic time step, and substantially reduces the noise compared to measuring point values.  All quantities reported in this manner are output one thousand times per binary orbit.

\section{Single-object disks} \label{sec:single}
Before studying circumbinary disks, it is useful to examine the steady-state profiles of accretion disks around single objects. Such simulations allow us to verify our torque-free sink implementation, and to more easily visualize the impact of various tunable parameters. Thus, we investigate the evolution of constant-$\nu$ accretion disks to a steady state, using a variety of sink parameters, and to explore the effects of different smoothing lengths. 

We set $r_s=0.1$ and explore $\epsilon_g=\{0.03, 0.05, 0.1\}$ for each of $\gamma=\{4, 8, 16\}$, using both Plummer and spline softening schemes, and both standard and torque-free sinks, for a total of 36 simulations. All simulations used $\nu=0.01$. 

The grid extended from $r=0$ to $r=10$, using 128 cells, linearly spaced for $r<0.4$ and logarithmically spaced for $r\geq 0.4$. Although the initial conditions were axisymmetric, we ran simulations using the full $r-\phi$ plane to catch potential errors or instabilities related to the sink prescriptions. All simulations used a density floor of $10^{-3}$.

The initial surface density profile was given by
\begin{equation}
\Sigma(r)=\exp\left[ -\left(r/r_e \right)^{-\xi} \right],
\end{equation}
where we set $r_e=0.15$ and $\xi=10$, setting up a cavity to be filled by accreting gas.
The initial angular velocity profile was given by 
\begin{equation}
\Omega^2 = \Omega_k^2(\mathcal{R}) + \frac{1}{\mathcal{R}\Sigma}\frac{dP}{dr}
\end{equation}
to satisfy centrifugal balance, where $\Omega_k$ is the the Keplerian angular velocity at a given radius, and $\mathcal{R}$ is a radial coordinate softened according the gravitational softening scheme used in each simulation. The radial velocity was set initially to the viscous rate,
\begin{equation}
v_r = -\frac{3\nu}{2\mathcal{R}}.
\end{equation}

\begin{figure}[ht]
\centering
\includegraphics[width=\columnwidth]{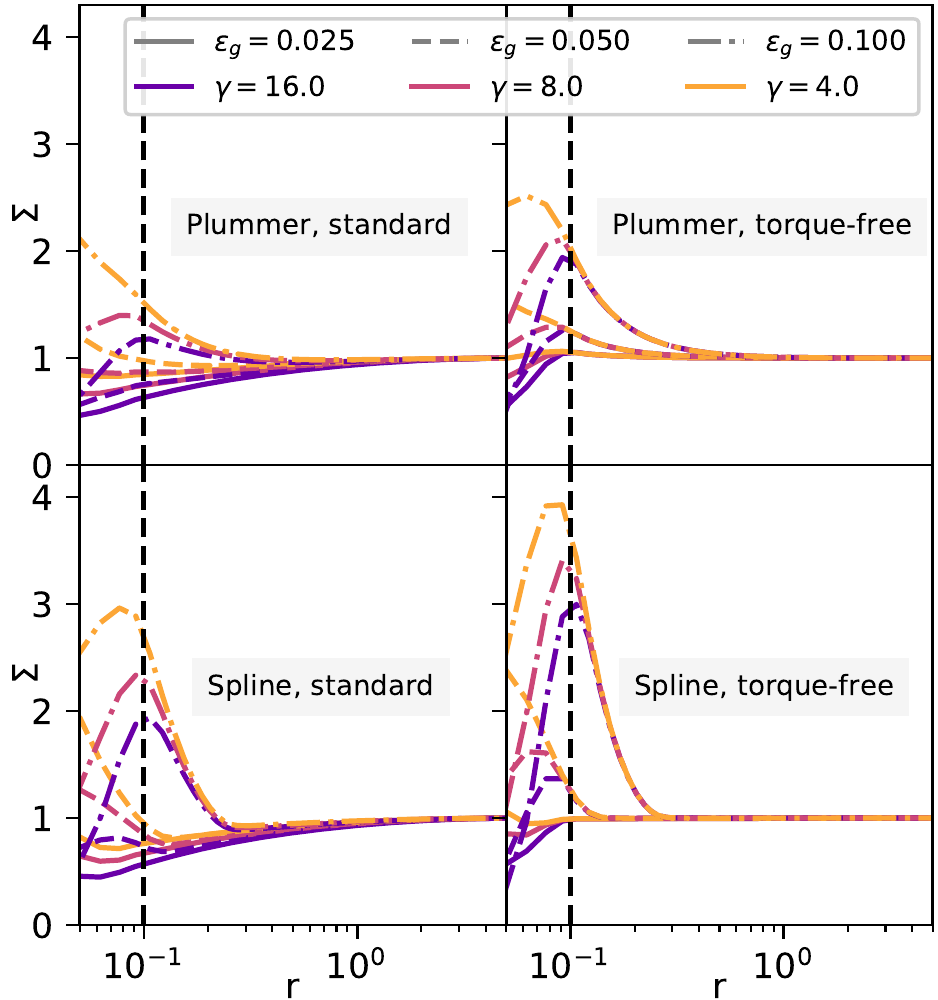}
\caption{1D steady state accretion disk surface density profiles. Line color indicates sink rate ($\gamma$), while line style indicates gravitational softening length ($\epsilon_g$). Vertical black dashed lines indicate the sink radius $r_s$.}
\label{fig:singleDisks}
\end{figure}

Each simulation was run until time $t=30\pi$. The viscous timescale at the cavity edge is on the order  $t_\nu(r_e)\sim \frac{2}{3}r_e^2/\nu\sim 1.5$, so the disks have more than enough time to settle into a steady state.  In an exactly Newtonian potential, absent source terms, the steady-state surface density profile is one of constant $\Sigma$. With a softened gravitational potential, the equilibrium density profile increases within $ r \lesssim \text{few}\times \epsilon_g$.  We present final surface density profiles in Figure \ref{fig:singleDisks}.

Figure \ref{fig:singleDisks} demonstrates the main result: for standard sinks (i.e. Equation \eqref{eq:standardMomSink}) the equilibrium surface density profile depends on the sink rate $\gamma$, even well outside the sink radius. Indeed using standard sink terms, regardless of softening scheme, surface density profiles for a given softening length still depend on $\gamma$ at more than ten times $r_s$. However, when using torque-free sinks (i.e. Equation \eqref{eq:tf1}) the surface density profiles have minimal dependence on $\gamma$ outside of $r_s$. Furthermore, any $\gamma$ dependence occurs within 1 to 2 cells of $r_s$, which is reasonable given that \Disco{} is spatially 2nd-order. 

 For the simulations that used torque-free sinks, such that the density profiles outside of the sinks are independent of $\gamma$, the effects of the softened potential on the surface density profile are more easily visible. As expected, since the spline-softened potential matches the Newtonian potential exactly for $r>2.8\epsilon_g$, so does the surface density. The Plummer sphere potential on the other hand leads to noticeable differences in the surface density profile significantly further from the sink. However, the spline potential leads to higher peaks in the surface density profile compared to a Plummer sphere potential with the same minimum. 

\section{Circumbinary Disks}

We have carried out a suite of circumbinary disk simulations, considering only circular binaries with semi-major axis $a$, but varying mass ratios $q=M_2/M_1$ where $M_1$ is the mass of the primary and $M_2$ is the mass of the secondary. We survey $q=\{1.0, 0.8, 0.6, 0.3, 0.2, 0.1\}$, testing both standard and torque-free sinks for each. Our default parameters were $\gamma=3.0$, $\epsilon_g = 0.033a$, $r_s=0.033a$ with Plummer softening, and $\mathcal{M}=10$ with a locally isothermal equation of state \eqref{lociso} for all simulations. We performed a larger suite of tests at $q=0.3$ and $q=1.0$, varying the sink rate $\gamma=\{1.0, 3.0, 10.0\}$, using Spline softening, and at $q=1.0$ with a simplified implementation of the viscous stress tensor (see Appendix \ref{visc}). We ran each simulation for 2000 binary orbital periods, which corresponds to a viscous timescale at $r\sim 4$, and is a common choice in the literature \citep[e.g.][]{{2020arXiv201009707Z},{2020ApJ...900...43T}}.

Our simulation grid extended from $r=0$ to $r=30a$, with cells spaced linearly in r for $r<a$, and logarithmically in r for $r \geq a$, along with an outer boundary condition held fixed at the initial value, a similar grid setup to that of \citep{2020ApJ...901...25D}. We performed initial tests with spatial resolutions of 512, 768, and 1024 radial zones at mass ratios $q=\{1.0, 0.3, 0.1\}$, finding that for $q=0.3$, 768 zones were sufficient for convergence. On the other hand, we found that for $q=0.1$, increasing resolution led to a noticeable difference on the gravitational torque on the secondary. Thus, for most mass ratios we use 768 radial zones, but for q=0.1 we present results from simulations using 1024 radial zones. When changing resolution, we scale $r_s$ and $\epsilon_g$ linearly with the grid spacing. The grid spacing in $\phi$ was chosen to keep the cell aspect ratio as close to 1 as possible. Note that with 768 radial zones, our smallest cell size is $0.00574a$.

The initial surface density profile was given by
\begin{equation}
\Sigma(r)=\Sigma_0\exp\left[ -\left(r/r_e \right)^{-\xi} \right],
\end{equation}
where we set the $r_e=2.5$ and $\xi=30$, setting up a cavity initially. 
The initial angular velocity profile was given by 
\begin{equation}
\Omega^2 = \Omega_k^2(\mathcal{R})\left(1 + \frac{3a^2}{4\mathcal{R}^2}\frac{q}{(1+q)^2} \right) +  \frac{1}{\mathcal{R}\Sigma}\frac{dP}{dr},
\end{equation}
where we modify the angular velocity by both pressure gradients and the orbit-averaged quadrupole moment of a circular binary, and $\mathcal{R}$ is $r$ with a small constant added. Note that the initial angular velocity inside the cavity is rapidly erased by the binary. The radial velocity was set initially to the viscous rate,
\begin{equation}
v_r = -\frac{3\nu}{2\mathcal{R}}.
\end{equation}
The accretion rate at the boundary is thus $\dot{M}_0=3\pi\nu\Sigma_0$.

\begin{figure*}[ht]
\centering
\includegraphics{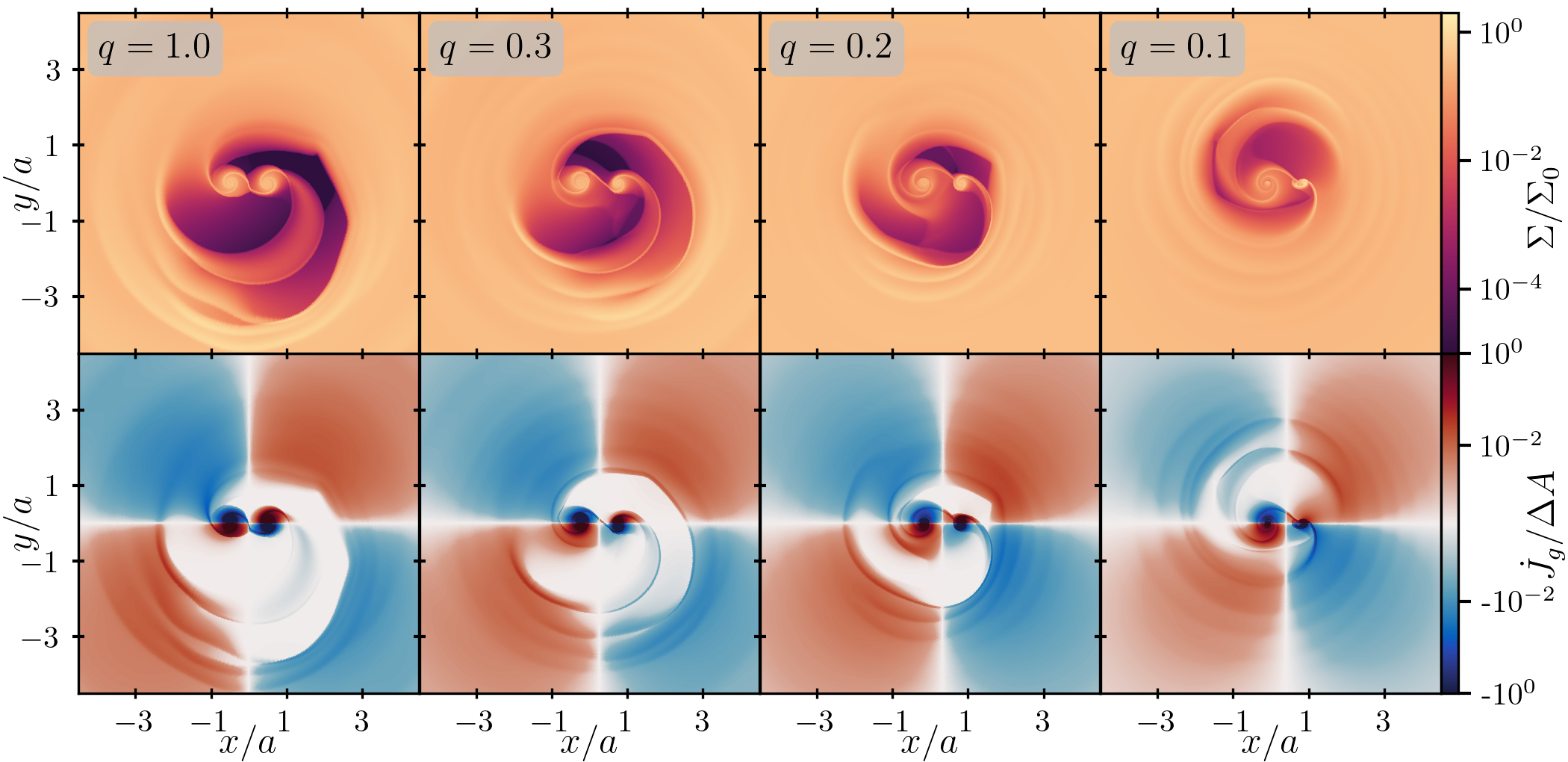}
\caption{Simulation snapshots after 1980 orbital periods. Top row: surface density profiles, normalized by $\Sigma_0$. Bottom row: gravitational torque density profiles. Mass ratios $q=\{1.0, 0.3, 0.2, 0.1\}$ are chosen to illustrate the change in cavity morphology near $q\sim0.2$. Each simulation shown used torque-free sinks and our default sink parameters.}\label{fig:qprofiles}
\end{figure*}

The general sequence of each simulation is as follows: gas flows into the initial cavity, forming accretion disks (``minidisks'')  around each component, and exciting density waves throughout the disk. 
An eccentric cavity forms around the binary, which precesses due to the gravitational quadrupole moment \citep{2012ApJ...749..118S, 2017MNRAS.466.1170M, 2020ApJ...905..106M}. Figure \ref{fig:qprofiles} shows snapshots of the evolved surface density for several mass ratios.

\begin{figure}
\centering
\includegraphics[width=\columnwidth]{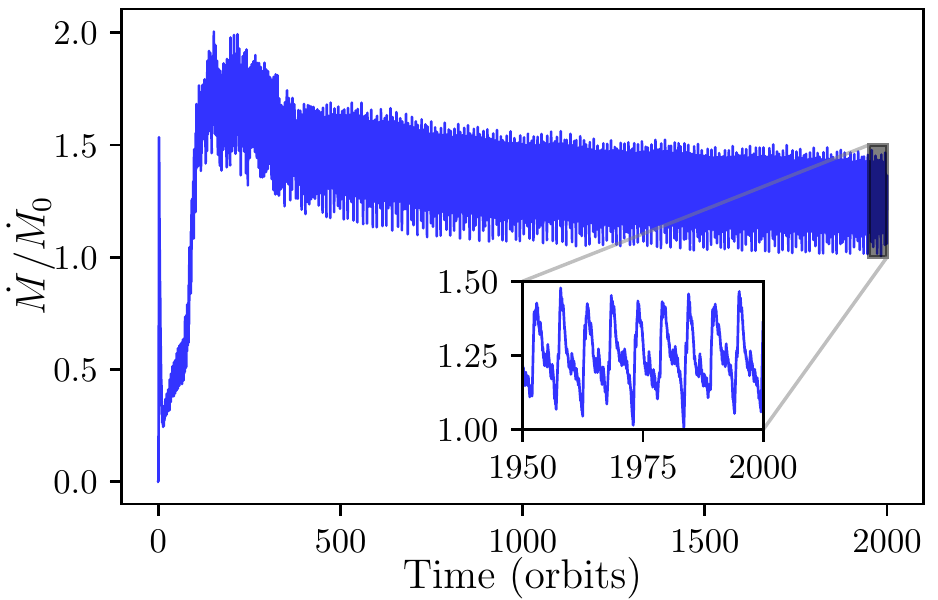}
\caption{A time series of the total accretion rate onto the binary over 2000 orbits of our fiducial $q=1.0$ simulation, using torque-free sinks. The inset shows the accretion rate over the final 50 orbits. The accretion rates are integrated over 0.02 orbits before being plotted.}
\label{fig:timeseries}
\end{figure}

It can take several hundred binary orbits for the basic flow patterns to be established, mindisks to form, and cavity to become eccentric.  During this time the accretion rate onto the binary is highly variable. After this point the binary accretes more steadily with a slow secular drift, becoming approximately steady (on the timescales considered here) after roughly $\sim1000-1500$ orbits, depending on the mass ratio. One example is shown in Figure \ref{fig:timeseries} for $q=1$, using torque-free sinks. On average, $\dot{M}/\dot{M}_0>1$, so although the accretion onto the binary settles into quasi-steady state, the disk is not in a global steady state, which would not be expected until a viscous timescale at the outer boundary of the domain. \citet{2020ApJ...889..114M} have compared circumbinary disk simulations of both finite and infinite disks. Disks with finite extent viciously spread and are precluded from reaching a true steady state. However, \citet{2020ApJ...889..114M} found that the two types of simulations were in agreement as long as results were normalized by the accretion rate $\dot{a}/\dot{M}$ instead of $\dot{a}$. Thus, even though the disks in our simulations are not in a global steady state, our inferences of binary evolution should be unbiased as long as the disk has settled into a quasi-steady state after the initial transient.

Accretion onto the binary occurs through streams which fall off the inner edge of the cavity wall, and are strongest when the outer binary component passes the eccentric cavity's periapse, modulating the accretion rate at the binary orbital frequency.  For $q\gtrsim0.2$, an overdensity forms on the cavity wall, which further modulates the accretion rate at the orbital period of the cavity, driving variability on timescales longer than the orbital period of the binary. For $q\lesssim0.2$ the cavity tends to be smaller and less elliptical and the accretion correspondingly steadier.

Changes in the cavity morphology can be seen clearly in Figure \ref{fig:qprofiles}, which illustrates the surface density and gravitational torque density for a single snapshot late in a series of simulations at different mass ratios, using torque-free sinks.

\subsection{Minidisk Structure}\label{miniStructure}

\begin{figure}
\centering
\includegraphics[width=\columnwidth]{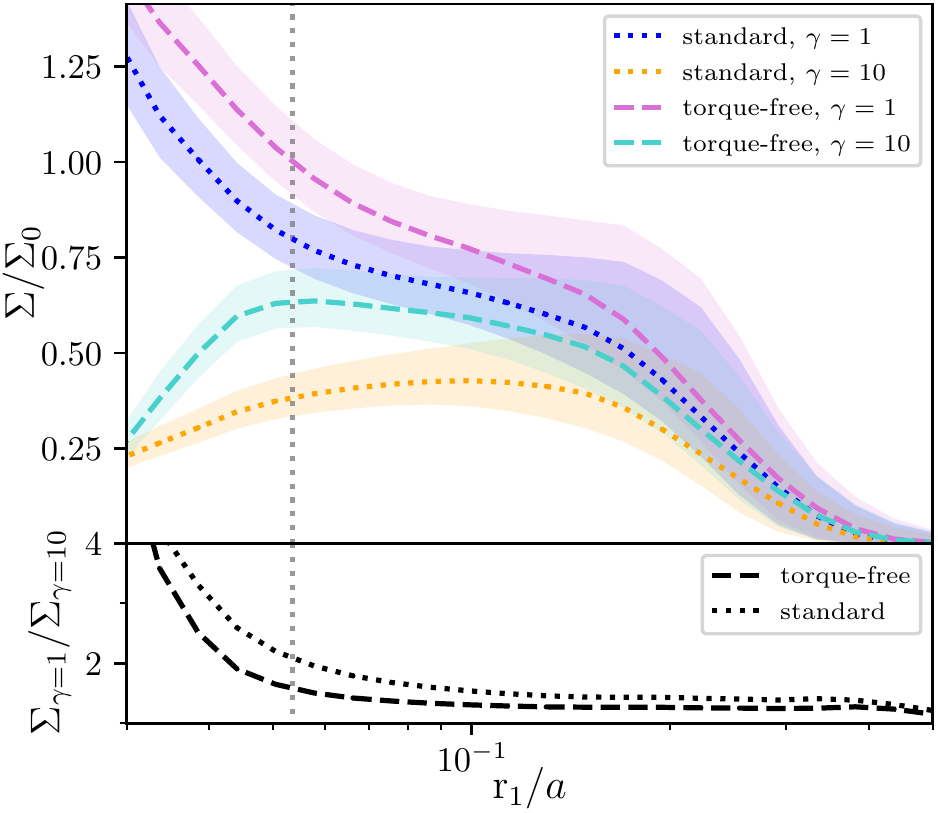}
\caption{Minidisk structures for $q=1.0$. The vertical dotted line indicates the radius at which $s_i=10^{-3}$ in Equation (\ref{GaussSquared}). Top panel: Minidisk surface density profiles. Lines indicate median values, while shaded regions correspond to the symmetric 68.3\% quantiles. In order from highest to lowest median surface density, the lines are: torque-free sinks, $\gamma=1$; standard sinks, $\gamma=1$; torque-free sinks, $\gamma=10$; and standard sinks, $\gamma=10$. Bottom panel: The ratio of the median surface density for the $\gamma=1$ and $\gamma=10$ runs for each sink type. The x-axis is the distance to the primary, but note that for $q=1$ the average minidisk profiles for the secondary are statistically identical to those shown here.}\label{fig:q1minis}
\end{figure}

\begin{figure}
\centering
\includegraphics[]{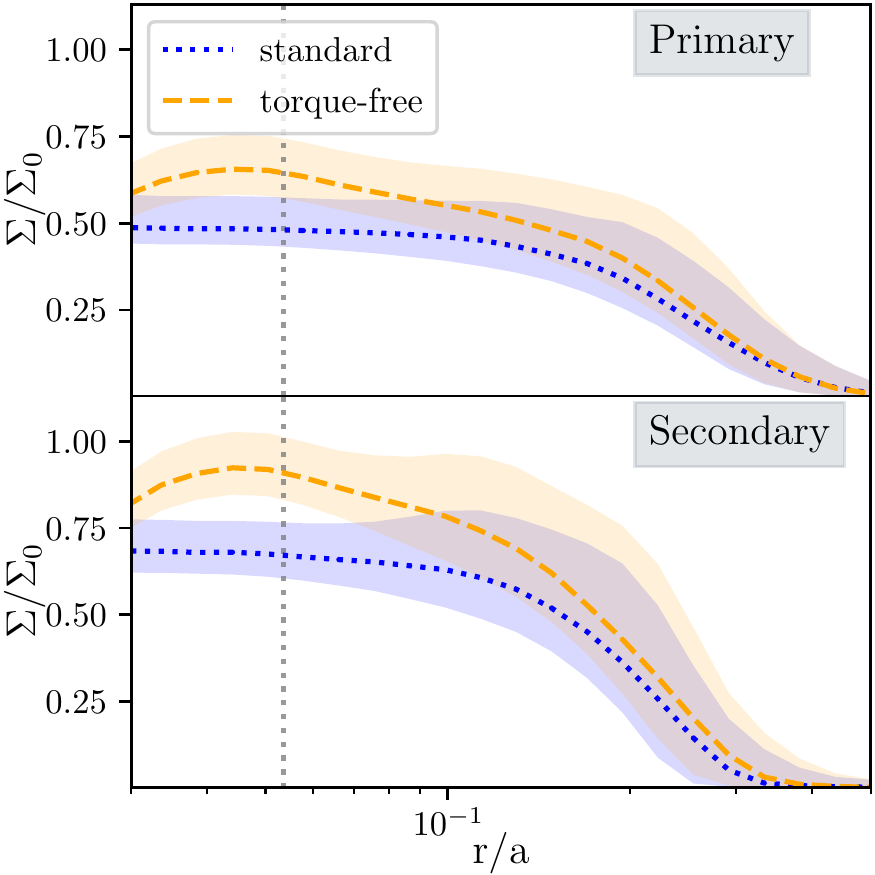}
\caption{Minidisk profiles for $q=0.6$. The vertical dotted line indicates the radius at which $s_i=10^{-3}$ in Equation (\ref{GaussSquared}). The top panel plots the minidisks for the primary, and the bottom panel shows the surface density of the secondary minidisk. Each minidisk is plotted as a function of the radial distance to its respective binary component. Blue line indicate the surface density from simulations using standard sinks, and orange lines indicate the surface density in simulation using torque-free sinks. The torque-free sinks lead to higher surface densities throughout most of the minidisks, and in the case of the secondary, a minidisk that extends further into the cavity.}\label{fig:q06minis}
\end{figure}

First, because the choice of sink has such a significant effect on the steady-state surface density profile in accretion disks with single objects, as demonstrated in Figure \ref{fig:singleDisks}, we examine the structure of the minidisks.  Unlike the steady single-object disks in Section \ref{sec:single},  minidisks are dynamic objects with global asymmetries and are subject to time variable gravitational forces and mass inflow \citep{2017ApJ...835..199R, 2017ApJ...838...42B}.  Minidisk structure also has important implications for the spectral appearance of accreting binaries, although we cannot properly capture this due to our assumption of an isothermal equation of state. 

We analyze minidisk structure by binning surface densities in the global $r-\phi$ grid by their distance to each object. We sample surface density values every orbit for the final 500 orbits of each simulation, and weight each sample by the volume of its cell. We then calculate the median values and symmetric 68.3\% quantiles of the surface density distribution function in each radial bin.  Since the minidisks are not axisymmetric, the 68.3\% quantiles capture the variability of the minidisk surface density in both time and azimuth.  Because these diagnostics are calculated in the frame of each object, we only consider matter within $a/2$ of each object. Because steady-state surface density profiles for disks around single objects are independent of sink rate for torque-free sinks but strongly dependent for standard sinks, we first focus on $q=1$ binaries, comparing runs with sink rates $\gamma=1$ and to those with $\gamma=10$. 

Surface density profiles for equal-mass binaries are presented in Figure \ref{fig:q1minis}. Across all mass ratios the sink prescription alters the structure of the minidisks.  Simulations with torque-free sinks consistently have higher-density minidisks, while simulations with faster sink rate $\gamma$ consistently have lower density minidisks.  Minidisks in simulations with torque-free sinks still show variation with sink-rate, unlike the steady-state disks in Section \ref{sec:single}, although the degree of variation is much reduced in comparison to simulations using standard sinks.  Figure \ref{fig:q1minis} shows that for $q=1$ binaries, the ratio of surface density between $\gamma=1$ and $\gamma=10$ for torque-free sinks is roughly two thirds the ratio for standard sinks.

The torque-free sinks are an improvement over standard sinks, in the sense that the resulting surface density profiles depend less on the sink rate. This will improve future studies of circumbinary disks by making predictions less dependent on free parameters such as the sink rate. 

For unequal mass ratio binaries, the symmetry between the minidisks is broken. Because of the weaker gravity of the secondary, its minidisk has a smaller extent compared to the primary. This is can be seen quite clearly for lower mass ratios in Figure \ref{fig:qprofiles}, as well as in Figure \ref{fig:q06minis} for $q=0.6$. Additionally, the secondary accretes preferentially from the cavity, which causes its minidisk to have a higher surface density. Together, these effects conspire to create sharp gradients in the surface density, necessitating a thorough treatment of viscous stresses as in Appendix \ref{visc}. Additionally, through interactions with the accretion streams and cavity wall, the size and mass of the minidisks can alter the global structure of the flow. 

Figure \ref{fig:q06minis} shows the minidisk surface density profiles for the $q=0.6$ case.  The effect of torque-free sinks is much more pronounced on the secondary, where the median density with standard sinks is significantly less than that in the torque-free run (outside the 15.9\% percentile of the torque-free profile in the bulk of the minidisk). The much more massive secondary minidisk corresponds to the cavity being slightly smaller, which leads to changes in the total angular momentum current through the disk (e.g. Figure \ref{fig:qseries})  and a slightly shorter dominant period in its variability (e.g. Figure \ref{fig:qVar}).

\subsection{Binary Evolution}

\begin{figure}[ht]
\centering
\includegraphics[width=.992\columnwidth]{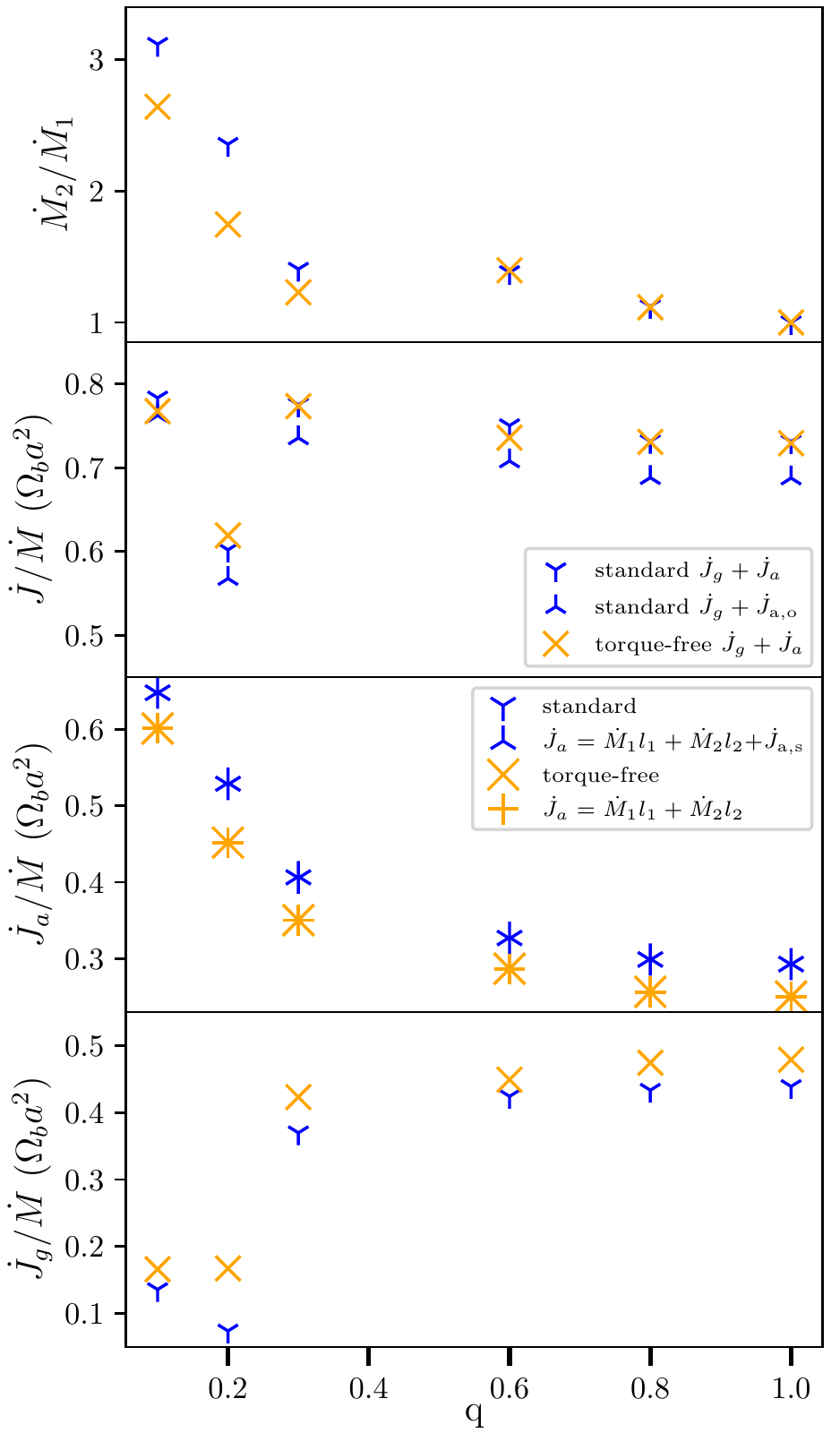}
\caption{Quantities significant for the evolution of the binary as a function of mass ratio. Blue symbols correspond to values from simulations using standard sinks, while orange symbols indicate values from simulations using torque-free sinks. Top panel: the ratio of accretion rates between the secondary and the primary. Second panel: total torque on the binary, with and without the spin component. Third panel: accretion torque on the binary. Bottom panel: gravitational torque on the binary.
}\label{fig:qseries}
\end{figure}

\begin{figure}[h]
\centering
\includegraphics[width=\columnwidth]{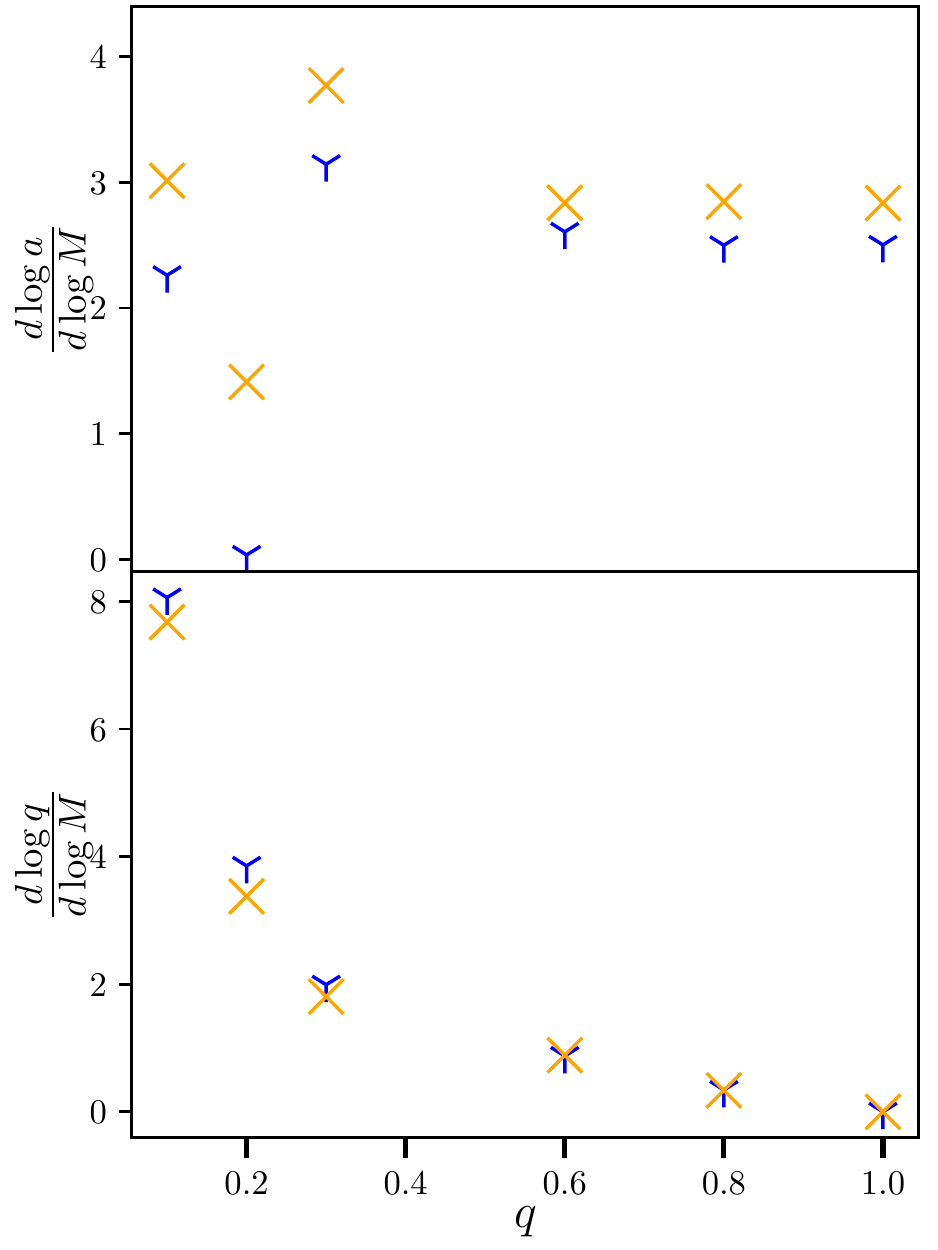}
\caption{Change in binary semi-major axis and mass ratio due to accretion as a function of mass ratio, given by Equations \eqref{eq:evolEqn3} and \eqref{eq:qevol}.}\label{fig:dadm}
\end{figure}

We now aim to understand the effect of the sink prescription on the secular evolution of binaries. We are primarily interested in the evolution of binary mass ratio $q$ and semi-major axis $a$. The eccentricity evolution is also of interest, although we only consider circular binaries in this work, and note that \citet{2020arXiv201009707Z} found that low-eccentricity binaries are circularized. We present the accretion rate ratio $\dot{M}_2/\dot{M}_1$ and torques in Figure \ref{fig:qseries}, and the change in semi-major axis, $d\log{a}/d\log{M}$, and the change in mass ratio, $d\log{q}/d\log{M}$, as functions of $q$ in Figure \ref{fig:dadm}.

Given the angular momentum of a binary
\begin{equation}
J_b=\frac{M_1M_2}{M}\sqrt{GMa\left(1-e_b^2\right)},
\end{equation}
where $e_b$ is the eccentricity of the binary and $M$ is the total mass, for a circular binary the torque on the binary can be expressed as 
\begin{equation} \label{eq:evolEqn1}
\frac{\dot{J}_{\rm orb}}{J_b} = \frac{\dot{M}_1}{M_1} + \frac{\dot{M}_2}{M_2} - \frac{1}{2}\frac{\dot{M}}{M} + \frac{1}{2}\frac{\dot{a}}{a},
\end{equation}
where $J_b=Ml_b$ is the angular momentum of the binary, 
$l_b = \Omega_ba^2q/(1+q)^2$ is the specific angular momentum of the binary.
When treating Equation \eqref{eq:evolEqn1} it is important to distinguish between $\dot{J}_{\rm orb}$ and $\dot{J}$ in order to retain only torques that influence the orbit. For torque-free sinks this is guaranteed by construction, but this point must be kept in mind when analyzing results from simulations using standard sinks.

We have found, as have a number of previous works \citep[e.g.][]{{2014ApJ...783..134F},{2020ApJ...901...25D},{2020ApJ...889..114M}}, that the lower-mass component of the binary accretes preferentially. 
Thus, we re-arrange equation \eqref{eq:evolEqn1} in the following manner: 
\begin{align}
    \frac{d\log{a}}{d\log{M}} &= 1+ 2\left[\frac{l_0}{l_b}-\frac{\dot{M}_1}{\dot{M}}(1+q)-     \frac{\dot{M}_2}{\dot{M}}\frac{1+q}{q}\right], \label{eq:evolEqn2} \\
        &= 2\left[\frac{l_0}{l_b}-\left(\frac{1+q}{1+\lambda}\right)\left(1+\frac{\lambda}{q}     \right)+\frac{1}{2}\right], \label{eq:evolEqn3}
\end{align}
where $l_0\equiv\dot{J}_{\rm orb}/\dot{M}$\footnote{This definition of $l_0$ is suitable for the evolution of the binary, but not for analyzing angular momentum transport through the disk. In the latter case, one should compare to $\dot{J}/\dot{M}$ rather than $\dot{J}_{\rm orb}/\dot{M}$} is the net angular momentum change in the binary per unit mass accreted and $\lambda\equiv\dot{M}_2/\dot{M}_1$ is the ratio of accretion rates.

Note that for equal-mass binaries, Equation \eqref{eq:evolEqn3} reduces to the condition that $l_0/l_b < 3/2$ is sufficient for a binary to shrink \citep{2017MNRAS.466.1170M}\footnote{The definition of $l_b$ used here differs by a factor of 4 from that in \citet{2017MNRAS.466.1170M}, but agrees with that in \citep{2020ApJ...900...43T}}

The evolution of the mass ratio $q = M_2 / M_1$ per unit mass accreted has a somewhat simpler expression:
\begin{equation}\label{eq:qevol}
    \frac{d \log q}{d \log M} = \frac{1+q}{1+\lambda}\left(\frac{\lambda}{q} - 1\right)\ ,
\end{equation}
which is positive (evolving towards equal masses) for $\lambda > q$ and negative (evolving towards unequal masses) for $\lambda < q$.

In Figure \ref{fig:qseries} we present results from our suite of simulations varying $q$. In agreement with previous results, we find that the secondary preferentially accretes, driving the binary towards an equal mass. For larger mass ratios ($q\gtrsim0.6$) the differences between torque-free and standard sinks are negligible, while for smaller mass ratios the primary receives a greater fraction of the total accretion rate when using torque-free sinks. Broadly, $\dot{M}_2/\dot{M}_1$ is a decreasing function of $q$, although the dependence is not monotonic. We note that \citet{2020ApJ...901...25D} also found non-monotonic features in $\lambda(q)$. However, our results suggest that previous studies using standard sink prescriptions may have overestimated the rate at which binaries approach equal mass by overestimating the rate of accretion onto the secondary. 

Torque-free sinks make only a small difference to the \emph{total} torque $\dot{J} / \dot{M}$ on the binary, as seen in the second panel of Figure \ref{fig:qseries}. At mass ratios $q=\{0.3, 0.8, 1.0\}$ the difference in torque is almost negligible, while at mass ratios $q=\{0.1, 0.2, 0.6\}$ there is a difference on the order of $2-3\%$.  However, in the case of the standard sinks the total torque includes a contribution from the spin torque, which will have no effect on the binary orbital evolution.  When considering only the torques on the \emph{orbital} angular momentum, the standard sinks consistently lead to lower values than the torque-free sinks by $\sim 4-9\%$ (with the possible exception of $q=0.1$).
This demonstrates that even when the sink prescription has little to no impact on global angular momentum transport through the circumbinary disk, it can have a substantial effect on how angular momentum is apportioned between orbit and spin.

The effect of the sinks on global angular momentum transport for some mass ratios is due to their ability to alter the structure of the mindisks, 
as seen in in Section \ref{miniStructure}, and which can in turn change the structure of the cavity. For $q=0.6$, the dependence of the total torque on the sink prescription is also reflected in differences in variability, as shown in Section \ref{sec:variability}, which stem from the cavity having a slightly different size. However, even when there is little change in $\dot{J}/\dot{M}$, the sink choice leads to differences in $\dot{J}_{\rm orb}$, and thus the evolution of the binary over time.

The accretion torque $\dot{J}_a$ includes three potential contributions: accretion of linear momentum from gas with a large velocity relative to the sink, accretion of spin angular momentum, and the direct accretion of mass from gas with a small relative velocity.  The third panel of Figure \ref{fig:qseries} demonstrates that for torque-free sinks the dominant contribution to $\dot{J}_a$ is from the direct accretion of mass, from which the torque is:
\begin{equation}
    \dot{J}_a = \dot{M}_1l_1 + \dot{M}_2l_2\ ,
\end{equation}
where $l_1=\Omega a^2 q^2/(1+q)^2$ is the specific angular momentum of the primary and $l_2=\Omega a^2/(1+q)^2$ is the specific angular momentum of the secondary. This is consistent with each \textit{component} of the binary on average accreting gas with specific angular momentum equal to its own.\footnote{This disagrees with the assumption that $\dot{J}_a=(\dot{M}_1 + \dot{M}_2)l_b$ made in \citet{2020ApJ...901...25D}, which did not directly record $\dot{J}_a$.}  The situation is similar for the standard sinks, except for an additional contribution from the spin torque $\dot{J}_s$.  This indicates that, for both sink prescriptions, the time-averaged contribution of direct accretion of linear momentum to the torque on the binary is negligible. Our results demonstrate that with torque-free sinks only recording $\dot{J}_g$, $\dot{M}_1,$ and $\dot{M}_2$ is sufficient to analyze the evolution of a circular binary, at least with sufficient numerical resolution.

Finally, the last panel of Figure \ref{fig:qseries} shows that the gravitational torque $\dot{J}_g$ can also be significantly affected by the choice of sink prescriptions.  Simulations with torque-free sinks consistently report larger gravitational torques on the binary, by more than a factor of two in the case of $q=0.2$.  It is not surprising that the gravitational torques can be sensitive to the sink prescription. The single-object disks in Figure \ref{fig:singleDisks} showed significant differences in the surface density profile between standard and torque-free sinks, which were also directly seen in the minidisk structure in Figures \ref{fig:q1minis} and \ref{fig:q06minis}.  \citet{2020ApJ...900...43T} showed that for $h/r \sim 0.1$ disks like those simulated here the gravitational torque is dominated by the minidisks.  Following those results, the \emph{gravitational} torques may become less sensitive to the sink prescription with thinner circumbinary disks.

For all mass ratios examined, regardless of sinks, we find $d\log{q}/d\log{M}>0$, in agreement with previous works. The trends in $d\log{q}/d\log{M}>0$, shown in the lower panel of Figure \ref{fig:dadm} clearly reflect those in $\lambda$ shown in Figure \ref{fig:qseries}, as expected from Equation \eqref{eq:qevol}: we find deviations due to differences in sink prescription at lower $q$, but little difference at larger $q$. 

At \emph{all} mass ratios we find significant differences in the evolution of the semi-major axis $d\log{a}/d\log{M}$ between torque-free and standard sinks, as shown in the first panel of Figure \ref{fig:dadm}. Our simulations with torque-free sinks lead to faster out-spiraling binaries.  The faster evolution is due to both the increased gravitational torque $\dot{J}_g$ and the decreased ratio of accretion rates $\dot{M}_2 / \dot{M}_1$ in simulations with torque-free sinks (shown respectively in the fourth and first panels of Figure \ref{fig:qseries}).  These effects add constructively, as seen in Equation \eqref{eq:evolEqn3}.  The difference is most dramatic for the $q=0.2$ binary, which is nearly stalled in the standard sink case but expands with $a \sim M^{1.5}$ when using torque-free sinks.

We list results for each simulation in our suite in Table \ref{tab:results} (in Appendix \ref{app:summary}), and in general find agreement in measured values of $l_0$, $d\log a/d\log M$, etc. with previous studies in cases with comparable mass ratios and Mach numbers \citep[e.g.][]{{2017MNRAS.466.1170M},{2019ApJ...871...84M},{2020ApJ...900...43T}} in our simulations using standard sinks.
Our results derived using torque-free sinks differ from those with standard sinks, especially in the accretion rate ratio $\dot{M}_2 / \dot{M}_1$ and binary orbital evolution $d \log a / d \log M$, and also display significantly less systematic variation with changes in sink rate $\gamma$.
We find that for either sink type, results using spline softening agree to a couple parts per thousand with those derived using Plummer softening. Sink rate on the other hand leads to significant differences in the measured quantities. For q=1, values of $d\log{a}/d\log{M}$ vary by $\pm 4\%$ for standard sinks, but by less than 0.1\% for the torque-free sinks, as a function of sink rate. Similarly, values of $\dot{J}_g$ measured from simulations using standard sinks vary by $\pm 3\%$ as a function of sink rate, but only by $\pm 0.2\%$ for simulations using torque-free sinks. Our simulations using $q=0.3$ follow the same trend, where the results from torque-free sinks have systematically less variance as a function of sink rate. 

\subsection{Variability}\label{sec:variability}
\begin{figure}
\centering
\includegraphics[width=\columnwidth]{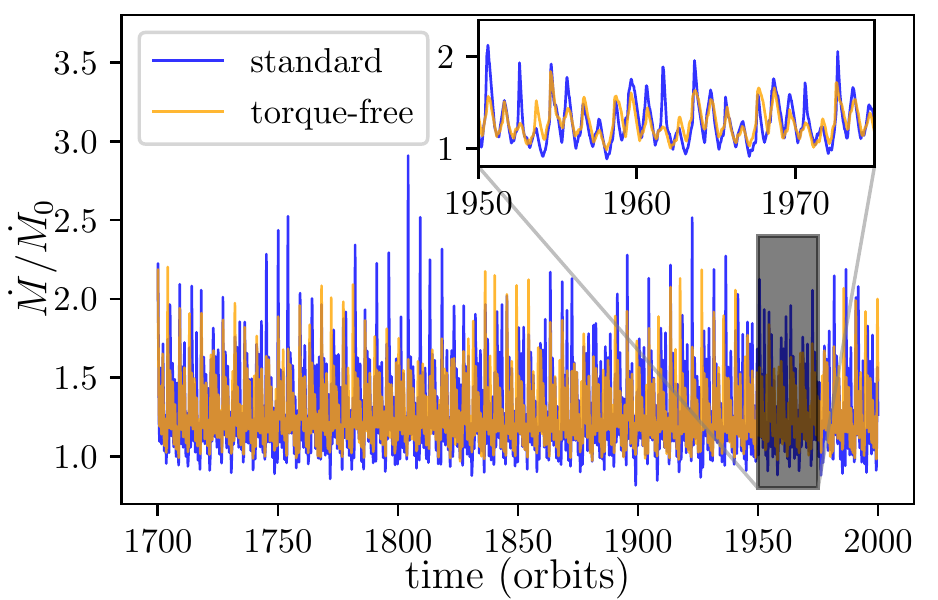}
\caption{A pair of accretion time series for the simulations with $q=0.3$ and $\gamma=10$, focusing on the last 300 orbits. The inset focuses on the accretion rates over 25 orbits. The accretion rates are integrated over 0.02 orbits before being plotted. Although the periodicity is similar between the two, the amplitude of accretion rate changes tends to be much larger for the simulation using standard sinks. }
\label{fig:timeseries2}
\end{figure}

Although electromagnetic emission cannot be inferred from isothermal simulations, the variability of the accretion flow, a standard proxy for electromagnetic luminosity \citep[e.g.][]{2014ApJ...783..134F,2020ApJ...901...25D} can be captured. We typically focus on variability in a statistical sense, although we include an example time series in Figure \ref{fig:timeseries2}. We begin by integrating our simulation outputs, originally taken 1000 time per orbit, down to a cadence of 50 times per orbit. We consider only data from the final 500 orbits of each simulation. We searched for variability on time scales ranging from $\sim0.1$ orbits to $~\sim 200$ orbits, although variability on timescales shorter than $\sim 1$ orbit is minor. Because samples were not spaced evenly in time, we identified the strength of variability at a given frequency by constructing Lomb-Scargle periodograms \citep{{1976Ap&SS..39..447L},{1982ApJ...263..835S},{2010ApJS..191..247T}}. For each periodogram, we first scale the accretion rate time series so that it has a mean of zero a variance of unity, and analyze the variation of each time series separately.

\begin{figure}
\centering
\includegraphics[width=\columnwidth]{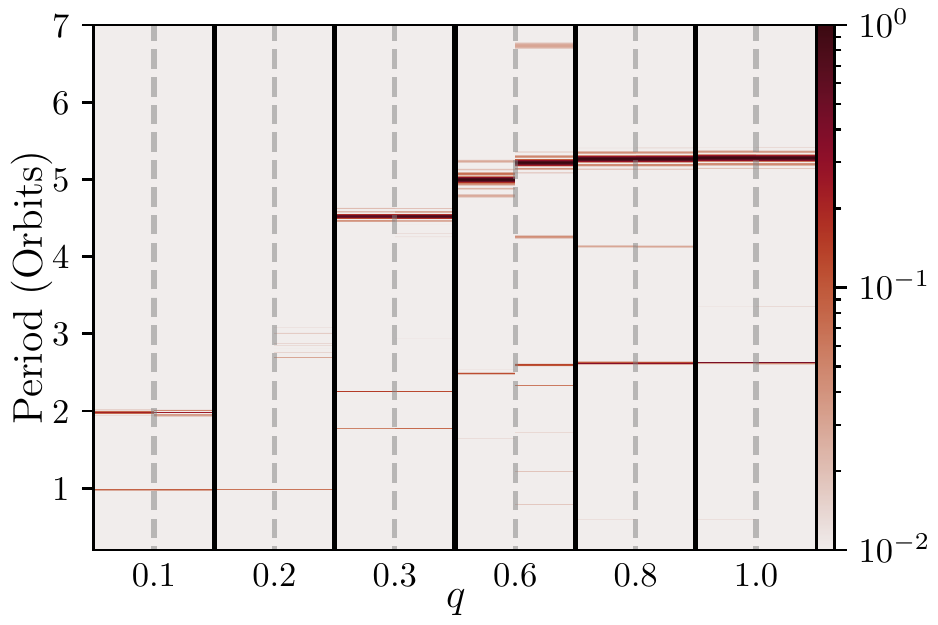}
\caption{Periodograms for our fiducial runs at each mass ratio. Different mass ratios are separated by solid black vertical lines. For each mass ratio, we display results for standard (left) and torque-free (right) sinks side-by side, separated by a dashed gray vertical line. Periodogram peaks are normalized by the largest amplitude across all mass ratios and sink types.}
\label{fig:qVar}
\end{figure}

In Figure \ref{fig:qVar},  we examine variability across mass ratios using our fiducial run parameters. We clearly see that for $q>0.2$, the variability largely occurs on time scales of $\sim 4.5-5.3$ orbits, although our resolution in $q$ is too coarse to precisely pinpoint this transition. We also see variability on harmonics of the dominant $\sim 4.5-5.3$ orbit variability for $q\geq0.3$, as well as variability on the period of the binary orbit, the latter of which is particularly strong for smaller mass ratios. This is especially clear in the case of the $q=0.2$ simulations, which almost exclusively show strong variability at the orbital period of the binary. The change in variability properties aligns cleanly with the transitions in gravitational torque reported in Figure \ref{fig:qseries} and the changes in cavity morphology displayed in Figure \ref{fig:qprofiles}. This $\sim5-$orbit accretion variability has been linked to the formation of a high-density clump, which we observe in our simulations for $q>0.2$, on the inner edge of the eccentric cavity, which then feeds the binary at the clump's periapse \citep[e.g.][]{2008ApJ...672...83M,2013MNRAS.436.2997D,2017MNRAS.466.1170M, 2019ApJ...879...76B}.

Typically, the dominant periods at which accretion varies are very similar between simulations using torque-free and standard sinks, although weaker peaks may vary, as can the widths of stronger peaks. The $q=0.6$ simulations deviate from this trend: runs with standard sinks show a faster dominant mode with $5.02$ orbit period compared to the $5.24$ orbit variability seen in the torque-free sink run.  This is likely due to a change in size of the cavity, which is fractionally larger in the torque-free run and would alter the orbital period of the lump on the cavity edge. Indeed we find the precession period of the cavity due to the binary quadrupole moment, which is sensitive to the cavity size \citep{2020ApJ...905..106M}, is also shorter for the standard sink $q=0.6$ run than the torque-free sink run.  The change in cavity size could plausibly be due to changes in the minidisk structure, which have a complex interaction with the cavity wall, although why this effect is only seen at $q=0.6$ is unknown.

\begin{deluxetable}{cccc}
\tablehead{
\colhead{$q$} & \colhead{$\gamma$} & \colhead{$c_v$ (standard sink)} & \colhead{$c_v$ (torque-free sink)}}
\startdata
1.0 & 1.0 & 0.215 & 0.192 \\
1.0 & 3.0 & 0.257 & 0.218 \\
1.0 & 10.0 & 0.318 & 0.253 \\
0.3 & 1.0 & 0.278 & 0.237 \\
0.3 & 3.0 & 0.421 & 0.330 \\
0.3 & 10.0 & 0.532 & 0.400 \\
\enddata
\caption{The normalized standard deviation of the accretion rate for various sink rates. }
\label{tab:rateVarTable}
\end{deluxetable}

Accretion onto the binary is clearly variable, and as demonstrated by Figure \ref{fig:qVar} the type of sink used can influence the global periodicity of the system and the range of frequencies at which accretion is periodic. We perform a simple test to measure the strength of variability in each simulation, assessing the normalized standard deviation (or coefficient of variation, $c_v$) of the accretion rate for a number of our simulations
\begin{equation}
c_v\left(\dot{M}\right) = \frac{\sqrt{\langle \dot{M}^2 \rangle - \langle \dot{M} \rangle^2}}{\langle\dot{M}\rangle},
\end{equation}
where $\langle.\rangle$ represents a time average. This lets us judge how consistent in amplitude the variability of the accretion rate is in time. As recorded in Table \ref{tab:rateVarTable}, we find that higher sink rates lead to larger-amplitude variability, regardless of sink type. However, when comparing results between torque-free and standard sinks, the increase in variation is less pronounced for torque-free sinks. This is consistent with our observation that the mass of the minidisks changes less as a function of sink rate when using torque-free sinks. We also find that at a given sink rate, the amplitude of  accretion rate variability is smaller for torque-free sinks, indicating that the global flow is steadier. This can be visualized clearly in the time series of accretion onto a pair of binaries. Figure \ref{fig:timeseries2} shows the total accretion rate over time for the $q=0.3$, $\gamma=10.0$ simulations. In accordance with the variations listed in Table \ref{tab:rateVarTable}, the accretion rate in the simulation using standard sinks has larger-amplitude variability, while the accretion rate in the simulation employing torque-free sinks remains closer to the mean.

\section{Conclusions}
Sink particles are an important sub-grid model in hydrodynamical simulations. We have found in simulations of circumbinary accretion that the sink prescriptions suggested by \citet{2020ApJ...892L..29D} significantly decrease 
the degree to which the evolution of binaries inferred from simulations depends on the sink rate.
Additionally, the sink prescription treats angular momentum in a manner more consistent with accretion onto an object much smaller than the grid scale, leading to minidisk surface density profiles that depend less strongly on sink rate. These different minidisks in turn can cause differences in the periodicity of accretion onto the binary, due to the coupling between the minidisks and cavity walls. We find the amplitude of accretion variability also depends on the sink rate, but to a lesser extent for torque-free sinks. Because of their improvements with regard to minidisk profiles and accretion variability, torque-free sinks will be vital for predicting observable features of supermassive black hole binaries and improve connection to observations.

Concerning the orbital evolution of binaries, consistent accounting of both accretion and gravitational torques, as well as the ratio of accretion rates, are of the utmost importance, as evidenced by the severe changes in $d\log{a}/d\log{M}$ shown in Figure \ref{fig:dadm}. The growth of binary separations as they accrete towards equal mass is particularly important with respect to binary stars, especially the population of near equal-mass binaries ($q\gtrsim0.95$) at wide separations ($a>1000~\rm{AU}$) reported by \citet{2019MNRAS.489.5822E}. At least in hotter ($h/r \sim 0.1$) disks such as those considered here, binaries with prograde circumbinary disks are expected to widen very rapidly as they increase in mass.

Studies of intermediate mass ratio inspirals have found that the surface density of gas within the Hill sphere of the secondary leads to a strong changes in the gravitational torque on the secondary due to the inclusion of a sink term over the course of the inspiral \citep{2021MNRAS.501.3540D}. Our results suggest that torque-free sinks have potential to improve such studies in the future, as they interfere less with the global flow pattern. 

Due to the sensitivities of the global flow on the sink rate, and how this sensitivity is reduced through the use of torque-free sinks, along with the fact that torque-free sinks are more accurate sub-grid models of accretion onto unresolved point masses, these sink treatments will be vital for comparing circumbinary disk simulations with all manner of observational data.

\section*{Software}

\Disco{} \citep{2016ApJS..226....2D}, \texttt{matplotlib} \citep{4160265}, \texttt{numpy} \citep{5725236}

\section*{Acknowledgments}
The authors are grateful to Paul Duffell, Ryan Westernacher-Schneider, and Cole Miller for many useful suggestions and discussions.
The Flatiron Institute is supported by the Simons Foundation. The numerical experiments presented in this paper were conducted in part on the Rusty Cluster at the Flatiron Institute and the Popeye-Simons System at the San Diego Supercomputing Center. The authors acknowledge the University of Maryland supercomputing resources (http://hpcc.umd.edu) that were made available for conducting the research reported in this paper; and the YORP cluster administered by the Center for Theory and Computation within the University of Maryland Department of Astronomy.

\appendix

\section{Viscosity implementation}\label{visc}
\noindent
We have improved the viscosity implementation in \Disco{} compared with \texttt{v1.0}, which made approximations which are not universally valid in a dynamic system such as circumbinary accretion.  In particular, \texttt{v1.0} made use of cancellations between the radial and angular momentum equations which are only valid in the constant dynamic viscosity limit.

The upgraded \Disco{} solves the compressible Navier-Stokes equations:
\begin{equation}
    \partial_t \left(\Sigma v_i\right) + \nabla_j\left( \Sigma v_i v^j + P\delta_i^j - 2\Sigma\nu \sigma_i^j\right) = -\Sigma \nabla_i \Phi  \ , \label{eq:momentum_components}
\end{equation}
    where $\nabla_i$ denotes the covariant derivative, superscripts $^i$ (subscripts $_i$) denote contravariant (covariant) tensor components in the coordinate basis, and we have neglected the momentum sink term for brevity.
    
    In curvilinear coordinates with metric tensor $g_{ij}$ any vectorial conservation law such as Equation \eqref{eq:momentum_components} takes the form: 
\begin{align}
    \partial_t u_i + \nabla_j f_i^j &= s_i &&\text{covariant derivatives} \\
    \partial_t u_i + \frac{1}{\sqrt{g}}\partial_j \left(\sqrt{g} f_i^j\right) &= s_i + \frac{1}{2}f^{jk}\partial_i g_{jk} &&\text{coordinate derivatives} \label{eq:conserved_vec_coord}
\end{align}
    
    In the above $u_i$ is the conserved vector quantity (e.g. the momentum), $f_i^j$ the flux tensor (e.g. the stress), and $s_i$ the vector of source terms (e.g. external forces).  Expressing the equations in terms of coordinate derivatives, necessary to incorporate into numerical algorithms, generates extra geometric source terms.
       
    The velocity shear tensor $\sigma_{ij}$ is symmetric, trace-free in $d$ dimensions, and defined by:
\begin{align}
    \sigma_{ij} = \frac{1}{2}\left(\nabla_i v_j + \nabla_j v_i\right) - \frac{1}{d}g_{ij} \nabla_k v^k\ .
\end{align}
    
    From this point we focus on polar coordinates $(r, \phi)$ in two dimensions with metric $g = \text{diag}(1, r^2)$, although we note our upgraded viscosity implementation in \Disco{} is written in three dimensions.  The divergence $\nabla_k v^k$ and shear tensor $\sigma_{ij}$ have values:
\begin{align}
    && \nabla_k v^k &= \partial_r v^r + \frac{1}{r}v^r + \partial_\phi v^\phi & & \\
    \sigma_{rr} &= \partial_r v^r - \frac{1}{d} \nabla_k v^k &  \sigma_{r\phi} = \sigma_{\phi r} &= \frac{1}{2}\left(r^2 \partial_r v^\phi + \partial_\phi v^r\right) & \sigma_{\phi\phi} &= r^2 \partial_\phi v^\phi + r v^r - \frac{1}{d} r^2 \nabla_k v^k\ .
\end{align}
    
    Plugging these into Equations \eqref{eq:momentum_components} and \eqref{eq:conserved_vec_coord} gives the full set of viscous fluxes and sources used by the upgraded \Disco{}:
\begin{align}
    {f_r^r}_{\viscNew} &= -2\Sigma \nu \left(\partial_r v^r - \frac{1}{3}\nabla_k v^k\right)                   & {f_\phi^r}_{\viscNew} &= -\Sigma \nu \left( r^2 \partial_r v^\phi + \partial_\phi v^r\right) \label{eq:fr_newvisc}\\
    {f_r^\phi}_{\viscNew} &= -\Sigma \nu \left(\frac{1}{r^2} \partial_\phi v^r + \partial_r v^\phi \right) & {f_\phi^\phi}_{\viscNew} &= -2\Sigma \nu \left(\partial_\phi v^\phi + \frac{1}{r} v^r - \frac{1}{3}\nabla_k v^k \right) \label{eq:fp_newvisc}\\
    {s_r}_{\viscNew} &= -2\Sigma \nu \frac{1}{r^2}\left(r\partial_\phi v^\phi + v^r - \frac{1}{3} r \nabla_k v^k \right) & {s_\phi}_{\viscNew} &= 0 \label{eq:s_newvisc}
\end{align}

    We set $d=3$ in the shear tensor, even in two dimensional simulations, to remain consistent with the microphysical definitions of shear and bulk viscosities \citep{LandauLifshitzFluids}.
    
    For comparison, Version \texttt{1.0} of \Disco{} used the following viscous fluxes and source terms, expressed here in the coordinate basis \citep{2016ApJS..226....2D}:
\begin{align}
    {f_r^r}_{\viscOrig} &= -\Sigma \nu \partial_r v^r                   & {f_\phi^r}_{\viscOrig} &= -\Sigma \nu r^2 \partial_r v^\phi \label{eq:fr_oldvisc}\\
    {f_r^\phi}_{\viscOrig} &= -\Sigma \nu \left(\frac{1}{r^2} \partial_\phi v^r - \frac{2}{r} v^\phi\right) & {f_\phi^\phi}_{\viscOrig} &= -\Sigma \nu \left(\partial_\phi v^\phi + \frac{2}{r} v^r\right) \label{eq:fp_oldvisc}\\
    {s_r}_{\viscOrig} &= -\Sigma \nu \frac{v^r}{r^2} & {s_\phi}_{\viscOrig} &= 0 \label{eq:s_oldvisc}
\end{align}
    
    These expressions were derived from the full set of fluxes assuming constant dynamic viscosity (ie. constant $\Sigma \nu$) and $d=2$, with a goal of eliminating gradients transverse to the direction of the flux.  They retain many salient features relevant for astrophysics, in particular the $\partial_r v^\phi$ term in $f_\phi^r$ which is the dominant term driving accretion in thin disks centered on the origin.  They also pass non-trivial code tests on constant density backgrounds, such as the Cartesian shear flow test performed in \citet{2016ApJS..226....2D}.
    
    Unfortunately, the assumptions made in deriving Equations (\ref{eq:fr_oldvisc}-\ref{eq:s_oldvisc}) are not globally valid in circumbinary accretion and other dynamic flows.  The inner cavity wall and outer minidisk edge both display strong density gradients where the incompressibility assumption breaks down.  Furthermore, the effective shear tensor used is not symmetric, which allows non-shearing flows (like rigid rotation) to generate erroneous viscous fluxes.
    
    The re-implementation of viscosity in \Disco{}, following Equations (\ref{eq:fr_newvisc}-\ref{eq:s_newvisc}), solves the full set of compressible Navier-Stokes equations with shear viscosity at the cost of requiring more gradient information when evaluating numerical fluxes and source terms.  \Disco{} automatically computes cell-centered, slope-limited gradients $\langle \partial_i v\rangle$ of all primitive variables (e.g. $\Sigma$, $v^r$, $v^\phi$, $P$) during the reconstruction phase of a time step \citep[see][for details]{2016ApJS..226....2D}.  The viscous source terms, Equation \eqref{eq:s_newvisc}, use these same gradients directly.  The viscous fluxes, Equations \eqref{eq:fr_newvisc} and \eqref{eq:fp_newvisc}, are evaluated at cell interfaces and require gradients from the cells on each side of the interface (left ``L'' and right ``R'') to be interpolated to the face center.  For gradients parallel to the face surface (transverse to the direction of the flux), the arithmetic mean of the gradients from the adjoining cells L and R is used.  That is, for a face with $x^i$-directed normal,  $\langle \partial_j v \rangle_{\mathrm{face}} = (\langle \partial_j v \rangle_{\mathrm{L}} + \langle \partial_j v \rangle_{\mathrm{R}}) / 2$ for $j\neq i$.  The gradient transverse to the face surface (parallel to the face normal) is computed directly via a second-order centered finite difference from the raw primitive values in the adjoing cells.  That is, for a face with $x^i$-directed normal, $\langle \partial_i v \rangle_{\mathrm{face}} = (v_{\mathrm{R}} - v_{\mathrm{L)}} / (x^i_{\mathrm{R}} - x^i_{\mathrm{L}})$, where $v_{\mathrm{L}/\mathrm{R}}$ are the cell centered values of $v$ in cells L and R, and $x^i_{\mathrm{L}/\mathrm{R}}$ are the $i$-component of the centroid positions of cells L and R, respectively.
    
\begin{figure}[h]
\centering
\includegraphics[width=0.5\columnwidth]{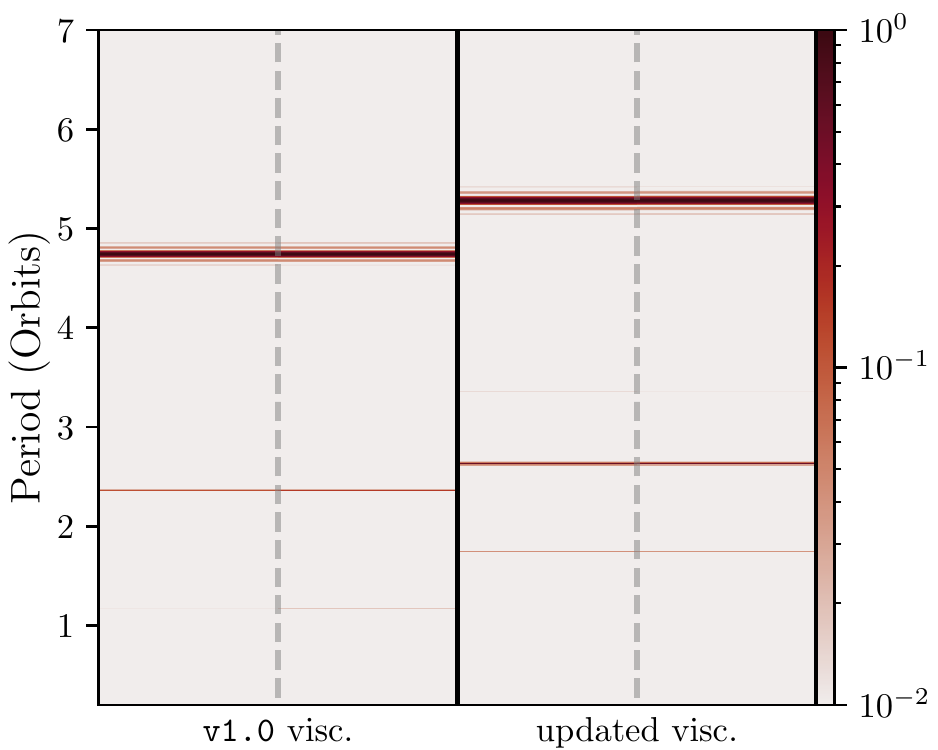}
\caption{Periodograms, normalized in the same manner as Figure \ref{fig:qVar}, for $q=1.0$. The solid black line divides the periodograms by viscosity implementation, and the gray dashed line separates the periodograms by the sink prescription used: the left column shows results standard sinks, and the right column shows the results using torque-free sinks. Between sinks, only the amplitudes of minor peiorogram peaks change for this mass ratio, but the viscosity prescription consistently alters the periodicity of accretion onto the binary.}
\label{fig:viscVar}
\end{figure}

    Figure \ref{fig:viscVar} shows the effect of the updated viscosity implementation on the accretion rate variability at mass ratio $q=1.0$.  The period of the dominant mode shifts considerably, indicating that the cavity is larger in the simulation using our updated viscosity prescription. The viscosity implementation used also dramatically alters the inferred properties of the evolution of the binary, as evidenced by the results in Table \ref{tab:results}. For example, $\dot{J}_g/\dot{M}$ changes by about $\sim10\%$ between the viscosity implementations, along with minor changes in $\dot{M}$. Because $\dot{J}_g/\dot{M}$ is lower in simulations using the \texttt{v1.0} viscosity implementation, the values of $d\log{a}/d\log{M}$ are correspondingly lower.

\section{Full Run Summary}
\label{app:summary}

\begin{deluxetable}{cccccccccccccc}
\tablehead{ \colhead{$q$} & \colhead{run} & \colhead{$\gamma$} & \colhead{softening} & \colhead{sink type} & \colhead{$\dot{M}$} & \colhead{$\dot{M}_2/\dot{M}_1$} & \colhead{$\dot{J}_g/\dot{M}$} & \colhead{$\dot{J}_a/\dot{M}$} & \colhead{$\dot{J}_s/\dot{M}$}&\colhead{$\dot{J}_{\rm orb}/\dot{M}$}& \colhead{$\dot{J}/\dot{M}$}&\colhead{$\frac{d\log a}{d\log M}$} & \colhead{$\frac{d\log q}{d\log M}$} }
\startdata
1.0 & $\gamma=1.0$&1.0&Plummer&standard&1.261 & 1.001 & 0.446 & 0.283 & 0.034 & 0.696 & 0.730 & 2.567 & 0.001 \\
1.0 & $\texttt{base}$&3.0&Plummer&standard&1.261 & 1.000 & 0.439 & 0.292 & 0.044 & 0.688 & 0.731 & 2.501 & 0.000 \\
1.0 & $\gamma=10.0$&10.0&Plummer&standard&1.262 & 1.000 & 0.426 & 0.306 & 0.058 & 0.674 & 0.732 & 2.389 & 0.000 \\
1.0 & $\texttt{spline}$&3.0&Spline&standard&1.263 & 1.000 & 0.438 & 0.292 & 0.044 & 0.687 & 0.730 & 2.494 & 0.000 \\
1.0 & $\texttt{v1 visc.}$&3.0&Plummer&standard&1.240 & 1.000 & 0.392 & 0.294 & 0.046 & 0.640 & 0.686 & 2.123 & 0.000 \\
1.0 & $\gamma=1.0$&1.0&Plummer&torque-free&1.261 & 0.998 & 0.478 & 0.250 & 0.000 & 0.728 & 0.728 & 2.823 & -0.002 \\
1.0 & $\texttt{base}$&3.0&Plummer&torque-free&1.261 & 1.000 & 0.479 & 0.250 & 0.000 & 0.729 & 0.729 & 2.834 & 0.000 \\
1.0 & $\gamma=10.0$&10.0&Plummer&torque-free&1.262 & 1.000 & 0.480 & 0.250 & 0.000 & 0.730 & 0.730 & 2.838 & 0.000 \\
1.0 & $\texttt{spline}$&3.0&Spline&torque-free&1.263 & 1.002 & 0.479 & 0.250 & 0.000 & 0.730 & 0.730 & 2.836 & 0.002 \\
1.0 & $\texttt{v1 visc.}$&3.0&Plummer&torque-free&1.241 & 1.000 & 0.435 & 0.251 & 0.000 & 0.686 & 0.686 & 2.488 & 0.000 \\ \hline
0.8 & $\texttt{base}$&3.0&Plummer&standard&1.261 & 1.122 & 0.433 & 0.298 & 0.043 & 0.688 & 0.732 & 2.498 & 0.342 \\
0.8 & $\texttt{base}$&3.0&Plummer&torque-free&1.260 & 1.115 & 0.474 & 0.257 & 0.000 & 0.731 & 0.731 & 2.845 & 0.335 \\ \hline
0.6 & $\texttt{base}$&3.0&Plummer&standard&1.268 & 1.380 & 0.424 & 0.326 & 0.042 & 0.708 & 0.750 & 2.605 & 0.874 \\
0.6 & $\texttt{base}$&3.0&Plummer&torque-free&1.263 & 1.397 & 0.449 & 0.287 & 0.000 & 0.736 & 0.736 & 2.836 & 0.887 \\ \hline
0.3 & $\gamma=1.0$&1.0&Plummer&standard&1.278 & 1.223 & 0.392 & 0.380 & 0.032 & 0.740 & 0.772 & 3.404 & 1.799 \\
0.3 & $\texttt{base}$&3.0&Plummer&standard&1.280 & 1.406 & 0.370 & 0.405 & 0.040 & 0.736 & 0.775 & 3.143 & 1.992 \\
0.3 & $\gamma=10.0$&10.0&Plummer&standard&1.283 & 1.559 & 0.351 & 0.430 & 0.052 & 0.729 & 0.781 & 2.913 & 2.132 \\
0.3 & $\texttt{spline}$&3.0&Spline&standard&1.279 & 1.392 & 0.371 & 0.404 & 0.040 & 0.735 & 0.775 & 3.154 & 1.978 \\
0.3 & $\gamma=1.0$&1.0&Plummer&torque-free&1.278 & 1.128 & 0.433 & 0.338 & 0.000 & 0.771 & 0.771 & 3.870 & 1.686 \\
0.3 & $\texttt{base}$&3.0&Plummer&torque-free&1.279 & 1.229 & 0.423 & 0.350 & 0.000 & 0.773 & 0.773 & 3.770 & 1.806 \\
0.3 & $\gamma=10.0$&10.0&Plummer&torque-free&1.280 & 1.317 & 0.414 & 0.359 & 0.000 & 0.774 & 0.774 & 3.671 & 1.902 \\
0.3 & $\texttt{spline}$&0.3&Spline&torque-free&1.278 & 1.226 & 0.423 & 0.350 & 0.000 & 0.773 & 0.773 & 3.766 & 1.803 \\ \hline
0.2 & $\texttt{base}$&3.0&Plummer&standard&1.192 & 2.357 & 0.074 & 0.528 & 0.034 & 0.568 & 0.602 & 0.035 & 3.855 \\
0.2 & $\texttt{base}$&3.0&Plummer&torque-free&1.201 & 1.746 & 0.167 & 0.452 & 0.000 & 0.619 & 0.619 & 1.411 & 3.378 \\ \hline
0.1 & $\texttt{base*}$&3.0&Plummer&standard&1.284 & 3.118 & 0.136 & 0.648 & 0.021 & 0.762 & 0.783 & 2.260 & 8.062 \\
0.1 & $\texttt{base*}$&3.0&Plummer&torque-free&1.275 & 2.642 & 0.166 & 0.601 & 0.000 & 0.768 & 0.768 & 3.012 & 7.677 \\
\enddata
\caption{Factors influencing the binary evolution, averaged over the last 500 binary periods of each simulation. Each simulation used $N_r=768,$ apart from those at $q=0.1$ which we have marked (*), and which used $N_r=1024$. Going left to right, beginning with column 6, we show: $\dot{M}$, the total mass accretion rate onto the binary, $\dot{M}_2 / \dot{M}_1$, the ratio of accretion rates between the secondary and primary, $\dot{J}_g / \dot{M}$, the gravitational torque on the binary normalized by the accretion rate, $\dot{J}_a / \dot{M}$ the accreted torque on the binary normalized by the accretion rate, $\dot{J}_s / \dot{M}$ the combined spin-torque on the binary normalized by the accretion rate, $\dot{J}_{\mathrm{orb}} / \dot{M}$, the orbital torque on the binary normalized by the accretion rate, $\dot{J} / \dot{M}$, the total torque on the binary normalized by the accretion rate, $d\log a/d\log M$, the growth rate of the binary separation per unit accreted mass, and $d\log q/d\log M$, the growth rate of the binary mass ration per unit accreted mass.} 
\label{tab:results}
\end{deluxetable}

\bibliographystyle{aasjournal}
\bibliography{references}
\end{document}